\parindent=0cm 

\font\gross=cmr10 scaled \magstep1

\font\mengen=bbm10

\font\csczwoelf=cmcsc10 scaled \magstep1

\font\ninebf=cmr7

\def\datum{\line{\hfil Cambridge, den \the\day.\the\month.\the\year}}
\def\date{\line{\hfil Berlin,  \the\month/\the\day/\the\year}}

\def\doppel{\magnification=\magstep1}

\def\titel#1#2{{\removelastskip\bigskip\goodbreak\noindent\mark{#1}\gross
#1\medskip \nobreak\noindent #2\unskip}}
\def\subtitel#1#2{{\removelastskip\bigskip\goodbreak\noindent\bf
#1\medskip \nobreak\noindent #2\unskip}}
\edef\ignore#1{}

\def\lrkopfzeile{\nopagenumbers\headline={\ifodd\pageno\ungeraderkopf\else
\geraderkopf\fi}
\voffset=2\baselineskip}
\def\geraderkopf{\rm\folio\ \dotfill\ \it\firstmark}
\def\ungeraderkopf{\it\firstmark\ \dotfill\ \rm\folio}

\def\part#1#2{{\partial #1\over\partial #2}}
\def\br#1#2{{#1\over #2}}
\def\frac#1#2{{#1\over #2}}

\def\sla#1{\hbox to 0cm{/\hss}#1}

\def\MR{\hbox{\mengen R}}
\def\MH{\hbox{\mengen H}}

\def\ID{\hbox{\mengen 1}}

\def\hence{\quad\Rightarrow\quad}

\def\lrvec#1{\vbox{\ialign{##\crcr$\leftrightarrow$\crcr\noalign{\kern-1pt\nointerlineskip}$\hfil\displaystyle{#1}\hfil$\crcr}}}
\def\pfeilrmittext#1{\setbox3=\hbox{#1\kern
0.5em}$\displaystyle\mathrel{\mathop {\hbox to
\wd3{\rightarrowfill}}^{\box3}}$}  
\def\pfeillmittext#1{\setbox3=\hbox{\kern
0.5em #1}$\displaystyle\mathrel{\mathop {\hbox to
\wd3{\leftarrowfill}}^{\box3}}$}  
\def\pfeilrmittextou#1#2{\setbox3=\hbox{#1\kern
0.5em}$\displaystyle\mathrel{\mathop {\hbox to
\wd3{\rightarrowfill}}^{\box3}_{\hbox{#2\kern 0.5em}}}$}  
\def\pfeillmittextou#1#2{\setbox3=\hbox{\kern
0.5em #1}$\displaystyle\mathrel{\mathop {\hbox to
\wd3{\leftarrowfill}}^{\box3}_{\hbox{\kern 0.5em #2}}}$}  
%

%

\def\putlogo{\vbox to 0cm{\hbox to
0cm{\noindent\line{\hss\epsfbox{AEIlogo.eps}}\hss}\vss}}

\newcount\figcount \figcount=0
\def\fig#1#2{\global\advance\figcount by1\midinsert\vskip #2
\centerline{Fig.\ \the\figcount : #1}\endinsert}
\def\figstuff#1#2{\global\advance\figcount by1\midinsert #2
\centerline{Fig.\ \the\figcount : #1}\endinsert}
\def\nextfig{Fig.~{\advance\figcount by 1\relax\the\figcount}}
\def\psfig#1#2{\global\advance\figcount
by 1 \midinsert\vbox{\centerline{\epsfbox{#2}}
\centerline{Fig.\ \the\figcount : #1}}\endinsert}
\def\rpsfig#1#2#3{\global\advance\figcount
by 1 \xdef\#3{\the\figcount}\midinsert\vbox{\centerline{\epsfbox{#2}}
\centerline{Fig.\ \the\figcount : #1}}\endinsert} 

\def\ez#1e#2{#1\cdot 10^{#2}}
\def\crossout#1{\hbox to 0cm{\raise 1ex \hbox{$\underline{\hbox{\phantom{#1}}}$}\hss}#1}
\def\foryoureyesonly#1{}
%
\def\paper#1#2#3#4#5#6#7#8#9{\item{\bf [#1]}#2: ``#3'', #4 {\bf
#5} (#6) p.\ #7 #8 \foryoureyesonly{#9}\par\goodbreak}
%
%

%
%

%
%

%
%
\newcount\chapterno
\newcount\subchapterno
\newcount\glno
\edef\rememberref#1{\relax}  
\def\rememberchapter#1{\relax}  
\def\remembersubchapter#1{\relax}  
\def\neueseitevorchapter{\vfill\supereject\ifodd\pageno\relax\else \noindent$
$\vfill\eject\fi} 
\def\kapitelnummer{\the\chapterno}
\def\chapter#1{\neueseitevorchapter\global\advance\chapterno by
  1\expandafter\rememberchapter{\the\chapterno . #1}\glno =0
  \subchapterno =0\titel{\the\chapterno . #1}} 
\def\subchapter#1{\global\advance\subchapterno by
    1\remembersubchapter{\the\chapterno .\the\subchapterno
    . #1}\subtitel{\the\chapterno .\the\subchapterno . #1}}
\def\gln#1{\global\advance\glno by 1\xdef#1{(\the\chapterno
.\the\glno)}\eqno {(\the\chapterno .\the\glno)\ifdraft\hbox to 0cm{\tt\string #1\hss}\else\relax\fi}}
\def\egln#1{\global\advance\glno by 1\xdef#1{(\the\chapterno
.\the\glno)}& {(\the\chapterno .\the\glno)\ifdraft\hbox to 0cm{\tt\string #1\hss}\else\relax\fi}}
\newif\ifdraft
\draftfalse
\def\draft{\drafttrue\def\vielluft{\bigskip}\def\neueseitevorchapter{\par}\def\foryoureyesonly##1{##1}\def\shlabel##1{\hbox
  to 0cm{\tt \expandafter\string ##1\hss}}}
\def\shlabel#1{\relax} 

\def\openbib{\def\aux{1}\openout\aux=\jobname.aux \immediate\write16{Putting
    references in \jobname.aux}
    \def\rememberref##1{\write\aux{cite(##1) on page \folio}}
    \def\rememberchapter##1{{\edef\x{\noexpand\write\aux{chapter[##1] on page
    \noexpand\folio}}\x}}
    \def\remembersubchapter##1{\write\aux{subchapter[##1] on page \folio}} }
\def\closebib{\closeout\aux \def\rememberref##1{\relax}}
\def\cite#1{\raise 0.5ex\hbox{\ninebf[#1]}\rememberref{#1}}

\newif\ifpdf
\ifx\pdfoutput\undefined\pdffalse\else\pdfoutput=1\pdftrue\fi

{\catcode`\%=11\catcode`\!=14
\gdef\bluebg{\pdfliteral{
}
\def\aeiseminar{\magnification = 2500\raggedright
        \ifpdf\pdfpagewidth=35true cm\hsize=
        \pdfpagewidth  
        \bluebg
        \BrickRed\else\textBrickRed\fi}
\def\color{\ifpdf\input pdfcolor\else\input colordvi\fi}



\doppel
\openbib
\def\neueseitevorchapter{\par}
\def\I{\hat\imath}
\def\J{\hat\jmath}
\def\K{\hat k}
\def\o#1{\sigma_#1}
{\nopagenumbers
\line{\hss DAMTP-2004-31}
\line{\hss hep-th/0403195}
\vskip 0.3\vsize plus 0.15 fil minus 3cm
\centerline{\csczwoelf Complexified moduli of non-commutative instantons} 
\vskip 4cm plus 0.4fil
\centerline{Robert C. Helling\footnote{$^1$}{{\tt helling@AtDotDe.de}}}
\centerline{DAMTP}
\centerline{Cambridge University}
\centerline{Wilberforce Road}
\centerline{Cambridge CB3 0WA}
\centerline{United Kingdom}
\vskip 2cm plus 0.4fil
{\bf Abstract:}

We revisit the generalised ADHM construction for instantons in
non-commutative space using a manifestly quaternionic formalism. This
leads to an identification of the self-dual part of $\theta^{\mu\nu}$
as the imaginary part of the size modulus of the instanton. 
\vskip 3cm plus 0.1fil
\eject}
\parskip=10pt
\chapter{Introduction}
In this note, we would like to demonstrate that the non-commutative
resolution of the instanton moduli space can be viewed as due to a
complexification of the size modulus. This achieved by reexpressing
the component construction of non-commutative instantons in the
quaternionic formalism of \cite{CSW} that mirrors the
realisation of instantons via the Hopf fibration of $S^7$. 

We find that the self-dual part of $\theta$, the measure of
non-commutativity, is an obstruction to solving the ADHM equation and
that one has to generalise the construction to include complexified
quaternions. In the end, however, the components of the physical gauge
fields are still real, but, as anticipated from other arguments, the
non-commutative instanton does not fit into a $SU(2)$ but only into a
$U(2)$.

The structure of this note is as follows: First we review the
quaternionic construction in the classical, commutative setting in
order to establish our notation. Furthermore, this allows us to highlight
the way different assumptions enter the construction and show where
the non-commutative generalisation differs. The next section discusses
this generalisation in depth and translates the quaternionic language
to the component language found in the existing literature on
non-commutative instantons.

\chapter{The commutative construction}
In a physically sensible gauge theory, the gauge field should have
real components as the imaginary parts of complex gauge fields would
have a kinetic term that would not be bounded from below. Usually,
this restriction is automatically dealt with by using a a compact
gauge group and a real formalism
in which no explicit factors of $i$ appear. However, in the case of
non-commutative coordinates, this is more difficult as commutators of
coordinates contain factors of $i$. In the simplest case of the
non-commutative plane that we will be concerned with in this note, the
basic commutator is given by
$$[x^\mu,x^\nu]=i\theta^{\mu\nu}.\gln\nc$$
Here $\theta^{\mu\nu}$ is a real, anti-symmetric matrix. As in quantum
mechanics, the factor of $i$ is needed as the coordinates should be
promoted to {\sl hermitian} operators and the commutator of such is
anti-hermitian.

In order to make reality (or anti-hermiticity of the connection)
explicit, we will re-express the non-commutative version\cite{NS} of the ADHM
construction\cite{ADHM} in the quaternionic formalism of \cite{CSW},
in which the anti-hermiticity is manifest.

Let us start by reviewing this construction in the case of commuting
coordinates. Later we will then contrast this with the
non-commutative version. The quaternionic formalism is based on
the identification
$$\MR^4\quad\hbox{(the space)} \quad\leftrightarrow\quad u(2)\quad \hbox{(the
gauge algebra)} 
\quad\leftrightarrow\quad \MH\quad \hbox{(the quaternions)}$$
where the first arrow relates two vector spaces but the second holds
not only in the Lie algebra (with commutators) but also in the
matrix (enveloping) algebra.

The quaternions $\MH$ are generated as a real algebra by $\o1$, $\o2$
and $\o3$ with the relations $\o1^2=\o2^2=\o3^2=-1$ and
$$\o1\o2=-\o2\o1=\o3 \qquad\hbox{and cyclic}.$$
It is convenient to define $\o4=1$ and then each $q\in\MH$ can be
written as $q=q^\mu\o\mu=q^4+q^1\o1+q^2\o2+q^3\o3$. Under
quaternionic conjugation, $q$ is mapped to $\bar
q=q^\mu\o\mu=q^4-q^1\o1-q^2\o2-q^3\o3$. The imaginary (or
anti-hermitian) are closed under taking commutators and will be
identified with $su(2)\subset u(2)$. Note that
$$\bar qq=q_1^2+q_2^2+q_3^2+q_4^2=\|q\|^2$$
is real and therefore commutes with all quaternions.

Now we can describe the commutative ADHM construction. For simplicity
we will restrict ourselves to the case of $su(2)=sp(1)$ gauge
theory. This starts with picking a $(k+1)\times k$ quaternionic matrix
$M$ that is linear in $x$, so that we can write it as 
$$M=B+Cx.$$ Note that $Cx$ is the product of a matrix with a
(quaternionic) scalar and thus is again a matrix rather than a vector
as a result of an inner product of a matrix and a vector! We have to
require that the matrix $R=M^\dagger M$ is in $GL(k,\MR)$ for all $x$,
that is, it is real and invertible. This requirement of reality is
what is usually referred to as ``the ADHM equations'' as it requires
the three imaginary parts of the matrix $R$ to vanish.

Next, we observe that the rank of $M$ can maximally be $k$, thus there
is a non-vanishing $x$-dependent vector $N\in\MH^{k+1}$ that is
annihilated by $M^\dagger$ (defined as the transpose of the
quaternionic conjugate of $M$):
$$M^\dagger N=0\gln\mn$$
Furthermore, we can require $N$ to be of unit length:
$$N^\dagger N=1\gln\nn$$
With this, we can define a gauge connection as
$A_\mu=N^\dagger\partial_\mu N$ that has a self-dual field-strength
with instanton number $k$.

Let us check these properties explicitly, so it becomes clear where
the various assumptions enter. First observe that the anti-hermiticity
of $A_\mu$ directly follows from taking $\partial_\mu$ of the normalisation
condition \nn. The field-strength is 
$$F_{\mu\nu}=\partial_\mu A_\nu +A_\mu A_\nu - \mu\leftrightarrow\nu
= (\partial_\mu N^\dagger)\partial_\nu N + N^\dagger(\partial_\mu N)
N^\dagger(\partial_\nu N) - \mu\leftrightarrow\nu.$$
Now, we use again the derivative of \nn\ to bring $F_\{\mu\nu$ to the
form 
$(\partial_\mu N^\dagger)(\ID_{k+1} - NN^\dagger)(\partial_\nu
N)$. From the definition of $N$ it follows that $\ID_{k+1} -
NN^\dagger$ is a projector on the subspace where $MM^\dagger$ is
invertible or, put differently, $MR^{-1}M^\dagger + NN^\dagger$ is the
unit matrix $\ID_{k+1}$. Using the derivative of
\mn, we arrive at
$$F_{\mu\nu} = (\partial_\mu N^\dagger)MR^{-1}M^\dagger(\partial_\nu
N) - \mu\leftrightarrow\nu = N^\dagger(\partial_\mu
M)R^{-1}(\partial_\nu M^\dagger )N - \mu\leftrightarrow\nu.$$
As $M$ is linear in $x$, we have $\partial_\mu M=C\o\mu$. Finally we
have to use the fact that $R$ is real and therefore commutes with
$\o\mu$ to arrive at
$$F_{\mu\nu}= N^\dagger CR^{-1}\o{{[\mu}}\bar\o{{\nu]}} C^\dagger N.$$
From the defining relation of the imaginary units we see
$$\o1\bar\o2=-\o3=-\o3\bar\o4$$
and cyclic which implies that $\o{{[\mu}}\bar\o{{\nu]}}$ is anti-self-dual
which concludes the proof of the anti-self-duality of $F$. Similarly,
$\bar\o{{[\mu}}\o{{\nu]}}$ is self-dual, and thus by baring the
above construction we would construct anti-instantons.
 
As a next step, let us do this construction explicitly for $k=1$, the
case of a single instanton. We start with
$$M=\pmatrix{b_1+c_1x\cr b_2+c_2x\cr}.$$ 
By shifting the origin in
$x$-space we can assume $b_2=0$. Thinking of $M$ as being a linear map
$M\colon\MH\to\MH^2$, we can choose an adapted basis of $\MH$ and
$\MH^2$ such that in this basis $c_1=0$ and $c_1=1$. After these
changes of coordinates
$$M=\pmatrix{b\cr x\cr}.$$
For this ansatz, we have to check the ADHM equations. We compute
$$M^\dagger M= \|b\|^2+\|x\|^2\gln\adhm$$
which is automatically real and everywhere invertible if
$b\ne0$. Next, we have to solve for $N=\pmatrix{n_1\cr n_2\cr}$. From
\mn, we find
$$0=\bar b n_1+\bar x n_2 \hence n_1=-{b\bar xn_2\over \|b\|^2}.$$
Similarly, \nn\ implies
$$\|n_2\|^2={\|b\|^2\over\|x\|^2+\|b\|^2}.$$
From this, we obtain
$$N={\|b\|\over\sqrt{\|x\|^2+\|b\|^2}}\pmatrix{-{b\bar
x\over\|b\|^2}\cr 1\cr} \lambda\gln\n$$
for some $x$ dependent $\lambda\in\MH$ with $\|\lambda\|=1$. Finally,
we can compute the connection:
$$A_\mu =
\bar\lambda{x_\mu+x\bar\o\mu\over\|x\|^2+\|b\|^2}\lambda+
\bar\lambda\partial_\mu\lambda,$$
from which we recognise $\lambda$ as a gauge parameter (unit norm is
equivalent to the restriction $su(2)\subset u(2)$) and up to a gauge
transformation 
$$A_\mu = {x_\mu+x\bar\o\mu\over\|x\|^2+\|b\|^2}.$$
After doing a rotation in
quaternion space we can assume $b\in\MR_{>0}$ and we recognise this as
the size modulus of the instanton. The four position moduli have been
fixed above by setting $b_2=0$. This concludes our review of the
quaternionic ADHM construction.

\chapter{The non-commutative ADHM-construction}
In the non-commutative case, we can try to apply the same strategy as
in the commutative case. There is only one important difference: Let us
compute $\bar xx$.
$$\bar xx=x^\mu x^\nu \bar\o\mu\o\nu=
x^{[\mu}x^{\nu]}\bar \o\mu\o\nu+x^\mu x^\nu
\bar\o{{(\mu}}\o{{\nu)}}.$$
For the first term, we use \nc\ and recall that the combination
$\bar\sigma\sigma$ is self-dual, while for the second we use
$\bar\o{{(\mu}}\o{{\nu)}}=\delta_{\mu\nu}$ which follows from the
definition of quaternions. So we arrive at
$$\bar xx=2i\theta^{\mu\nu}_{SD}\bar\o{{[\mu}}\o{{\nu]}}+\|x\|^2.$$
It turns out that this is in fact the only direct difference of the
non-commutative construction to the one of the previous section. 

As an immediate observation we recover the well known fact that only
the self-dual part of $\theta$ contributes. In fact, in the case
of an anti-self-dual background, none of the formulae change.

The second important observation is that the self-dual matrix
$\bar\o{{[\mu}}\o{{\nu]}}$ changes sign under quaternionic
conjugation. It is thus purely imaginary! This appears to pose an
obstruction to the ADHM equation \adhm\ if we try to construct a
non-commutative instanton:
$$M^\dagger
M=\|b\|^2+\|x\|^2+2i\theta^{\mu\nu}_{SD}\bar\o{{[\mu}}\o{{\nu]}}
\in^?GL(1,\MR)=\MR^*$$ 
Two of the three terms are manifestly real while the third is purely
imaginary (in the quaternionic sense that prevents it to commute with
other quaternions as discussed above; the $i$ does not play a role her)
and cannot be cancelled to yield a real sum. To our knowledge, this
obstruction has not been discussed in the literature on
non-commutative instantons, mainly due to the fact that it is manifest
only in the quaternionic formalism.

The reader might be worried\cite{BSST}\ that what we have described
here is really the construction of an instanton in $sp(2)$ rather than
$su(2)$ gauge theory and that although these theories are the same in
the commutative setting they might differ in non-commutative
space. Indeed, applying the rules\cite{CSW} for the ADHM construction for the
$su(N)$ series requires $M$ to be a quaternionic $4\times 1$ matrix
for which in addition to the above ADHM condition also
$$M^\dagger \o1 M$$
has to be real. However, it is not hard to check that up to conjugation in the
quaternions this additional condition is solved by
$$M=\pmatrix{b\cr\o2 b\cr x\cr \o2 x\cr}.$$
With this ansatz, the rest of the construction reduces to the case
discussed above.

The existing literature constructs instantons in non-commutative space
without encountering the problem of this imaginary contribution to the
ADHM construction.  Let us therefore translate our findings to the
conventional component formalism and see how the obstruction is
circumvented there. To be concrete, we will compare to the formalism
used in \cite{CKT}, but then making contact with other treatments of
the subject, for example
\cite{N}\cite{DN}\cite{KLY} should be straight forward. To do so we use
the following representation of the quaternionic units in terms of
$2\times 2$ matrices:
$$
\o1=\pmatrix{0&1\cr -1&0\cr}\quad
\o2=\pmatrix{0&i\cr i&0\cr}\quad
\o3=\pmatrix{i&0\cr 0&-i\cr}\quad
\o4=\pmatrix{1&0\cr 0&1\cr}$$
The reader should not be confused that although some of these matrices
have complex entries, we are still dealing with a real algebra, that
is the general element of $\MH$ is a linear combination of these
matrices with {\sl real} coefficients. For example, the quaternion
that we denote $x$ is in this matrix notation
$$x=\pmatrix{x^4+ix^3&x^1+ix^2\cr -x^1+ix^2& x^4-ix^3\cr}.$$
Furthermore, we choose coordinates that skew diagonalise
$\theta^{\mu\nu}$:
$$\theta^{\mu\nu} = \pmatrix{&\theta_1&&\cr -\theta_1&&&\cr
&&&\theta_2\cr &&-\theta_2&\cr}.$$
So the purely imaginary quaternion becomes
$$2i\theta^{\mu\nu}\bar\o\mu\o\nu = 4i (\theta_1+\theta_2)\o3=
\pmatrix{-4(\theta_1+\theta_2)&\cr &4(\theta_1+\theta_2)\cr}.\gln\problem$$
In order to fulfil the ADHM equations, this contribution should be
cancelled by a term from $b$. \cite{CKT} do this (in their equation
(6.11) adopted to our conventions) by taking $b$ to be the matrix
$$b=\pmatrix{&\rho\cr \sqrt{8(\theta_1+\theta_2)+\rho^2}&\cr}\gln\b$$
and indeed 
$$\eqalign{\bar bb&= \pmatrix{&\rho\cr
\sqrt{8(\theta_1+\theta_2)+\rho^2}&\cr}^\dagger \pmatrix{&\rho\cr
\sqrt{8(\theta_1+\theta_2)+\rho^2}&\cr} \cr
&=
\pmatrix{\rho^2+4(\theta_1+\theta_2)&\cr
&\rho^2+4(\theta_1+\theta_2)\cr} +  
\pmatrix{4(\theta_1+\theta_2)&\cr
&-4(\theta_1+\theta_2)\cr}.}$$
So, obviously $\bar xx+\bar bb$ is proportional to the unit matrix as
required by the ADHM equation and
thus real when translated back to quaternions.

How does this solution fit into the quaternionic framework? To answer
this question, we have to translate the ansatz \b\ by expressing it as
a linear combination of the $\o\mu$:
$$b=\left(\br\rho 2-\sqrt{2(\theta_1+\theta_2)+{\rho^2\over
4}}\right)\o1 -i\left(\br\rho
2+\sqrt{2(\theta_1+\theta_2)+{\rho^2\over 4}}\right)\o2 .$$
The important difference to what we tried above is that because of the
$i$ this is not a real linear combinations of the $\o\mu$ but an
element of the complexified quaternions! So, to solve the ADHM
equations in a self-dual background, we have to allow for complex
components. This together with the matrix inspired conjugation rule
$$\bar q= -\bar q^1\o1-\bar q^2\o2-\bar q^3\o3+\bar q^4$$
where the bar on the coefficients is complex conjugation makes $\bar
qq$ an arbitrary (ordinary, not complexified) quaternion:
$$\bar qq = \|q\|^2+2i \Im(\bar b^\mu
b^\nu)\bar\o{{[\mu}}\o{{\nu]}}$$
Once we have made the generalisation to complexified quaternions we
see that we could conjugate $b$ to a more conventional form and
parametrise the solutions of the ADHM equations as
$$b = \eta -\br i\eta \theta^{\mu\nu}\bar\o\mu\o\nu,\gln\bcomp$$
as this leads to
$$\bar bb =
\eta^2+{\theta_{SD}^2\over\eta^2}-2i\theta^{\mu\nu}\bar\o\mu\o\nu$$
and cancels the imaginary part of $\bar xx$.

In this expression we can interpret the real part as the square of the
size of the instanton (corresponding to $\|b\|^2$ in the commutative
case) and we recover the well known result that in the
presence of a non-commutativity with a self-dual component, the
minimal size of instantons is $\sqrt{2\theta_{SD}}>0$.

One might be worried that the complexification of the components of
$M$ and thus $N$ would complexify the components of the
physical field $A_\mu$ as well and thus thus render the
non-commutative instantons unphysical as negative kinetic terms arrise
for non-compact (for example complexified) gauge groups. However, the
effect of the complexification of $M$ is still physically acceptable:
Once again taking the derivative of $\nn$ (where now the adjoined
includes the conjugation on the complexified quaternions), one finds
$$A_\mu=N^\dagger\partial_\mu N$$ to be odd under conjugation (as in
the commutative case). If $A_\mu$ would be an ordinary quaternion this
would imply that its real part vanishes and $A_\mu$ is in $su(2)$. In
the complexified case however, it just means the coeffiecients of the
$su(2)$ generators in $A_\mu$ are real and the coefficient of the
$u(1)\subset u(2)$ is imaginary. But in our convention this means that
that $A_\mu$ is in the compact real form of $u(2)$. The effect of this 
generalization to complexified quaternions is just that the
non-commutative instantons do not fit in $su(2)$. One has to enlarge
the gauge group to $u(2)$, as it is well known for non-commutative
gauge theories.

However, the main result of applying the quaternionic formalism and
finding the need for the complexification is \bcomp, which shows that
what used to be the real size modulus in the commutative case has now
been complexified and $\theta_{SD}$, the self-dual component of the
background, appears as its imaginary part. This fits well into the
general pattern of how string theory avoids singularities in moduli
spaces: There is a real geometric modulus for which the theory becomes
singular at the origin. But string theory pairs this real modulus
together with the flux of a background field into a complex modulus so
the singularity at the origin can be avoided by turning on that flux
and thus going into the complex plane. 

\chapter{Discussion}
We translated the generalisation of the ADHM
construction of instantons to non-com\-mu\-ta\-tive spaces to the
quaternionic formalism. We found that
for $\theta_{SD}\ne 0$ it is not possible to solve the ADHM equations
over the quaternions. Rather we had to allow for quaternions with
complex coefficients.

From this point on, all expressions have complex coefficients in
principle, but we found that in the case of the gauge field $A_\mu$
which is a physical field as opposed to $M$ and $N$, anti-hermiticity
and thus compactness of the gauge group and positivity of the kinetic
term are all preserved. The only effect of the complexification is
that the trace will no longer vanish and thus the gauge field is in
$U(2)$ rather than $SU(2)$.

However, there remain a couple of open questions: In the commutative
case, we knew that $N$ is nowhere vanishing and thus could trivially
impose the normalisation \nn. For non-commutative $*$-multiplication
however, there is in general no $*$-division. Still one can calculate
inverses in the operator sense, at least for operators of the form
$1-O$, the geometric series gives inverses at least for operators not
having a spectrum which intersects the unit circle.

By the nature of these operator inverses, explicit solutions are not
available and thus we only have implicit knowledge of the physical
profile of non-commutative instantons.

Furthermore, for functions like $M$ and $N$ that depend on $x$, even
real expressions do not commute anymore and one has to be extremely
careful with orderings like in expressions of the form of \n.

\bigskip
Acknowledgements:

The author thanks Valentin Khoze for a helpful conversation and the
European Superstring Theory Network for financial support. This work
is partly supported by EU contract HPRN-CT-2000-00122

\chapter{References}
\bigskip
\parindent=2cm
\paper{ADHM}{Atiyah, M. F. and Hitchin, N. J. and Drinfeld, V. G. and                  Manin, Yu. I.}{Construction of instantons}{Phys. Lett.}{A65}{1978}{185--187}{{\tt } }{Cited: 2 }
\paper{BSST}{Bonora, L. and Schnabl, M. and Sheikh-Jabbari, M. M. and                  Tomasiello, A.}{Noncommutative SO(n) and Sp(n) gauge theories}{Nucl. Phys.}{B589}{2000}{461--474}{{\tt hep-th/0006091} }{Cited: 6 }
\paper{CKT}{Chu, Chong-Sun and Khoze, Valentin V. and Travaglini,                  Gabriele}{Notes on noncommutative instantons}{Nucl. Phys.}{B621}{2002}{101--130}{{\tt hep-th/0108007} }{Cited: 6 7 }
\paper{CSW}{Christ, Norman H. and Weinberg, Erick J. and Stanton, Nancy                  K.}{General self-dual Yang-Mills solutions}{Phys. Rev.}{D18}{1978}{2013}{{\tt } }{Cited: 2 2 6 }
\paper{DN}{Douglas, Michael R. and Nekrasov, Nikita A.}{Noncommutative field theory}{Rev. Mod. Phys.}{73}{2001}{977--1029}{{\tt hep-th/0106048} }{Cited: 6 }
\paper{KLY}{Kim, Keun-Young and Lee, Bum-Hoon and Yang, Hyun Seok}{Noncommutative instantons on R**2(NC) x R**2(C)}{Phys. Lett.}{B523}{2001}{357--366}{{\tt hep-th/0109121} }{Cited: 6 }
\paper{N}{Nekrasov, Nikita A.}{Noncommutative instantons revisited}{CMP}{241}{2003}{143--160}{{\tt hep-th/0010017} }{Cited: 6 }
\paper{NS}{Nekrasov, Nikita and Schwarz, Albert}{Instantons on noncommutative R**4 and (2,0) superconformal                  six  dimensional theory}{CMP}{198}{1998}{689--703}{{\tt hep-th/9802068} }{Cited: 2 }

\closebib

\bye